\documentstyle[12pt,german]{article}
\begin{document}
\hfill LMU 01/98\\
\\
\noindent
{\bf QUARK MASS HIERARCHY AND FLAVOR MIXING}\\
\\
\\
\\
\hspace*{2.5cm} Harald Fritzsch\\
\\
\hspace*{2.5cm} Sektion Physik, Universit"at M"unchen,\\
\hspace*{2.5cm} D--80333 M"unchen, Germany\\
\\
\\
\\
{\bf ABSTRACT}\\
\hspace*{0.3cm}
In view of the observed strong hierarchy of the quark and
lepton masses and
of the flavor mixing angles it is argued that the description of flavor
mixing must take this into account. One particular interesting way to describe
the flavor mixing, which, however, is not the one used today, emerges, which
is best suited for models of quark mass matrices based on flavor
symmetries. We conclude that the unitarity triangle important for $B$
physics should be close to or identical to a rectangular triangle. $CP$
violation is maximal in this sense.\\
\\
\\
\begin{center}
{\it Invited talk given at the}\\
{\it International Conference on Orbis Scientiae 1997:}\\
{\it Physics of Mass}\\ 
{\it Miami Beach, Florida (December 12--15, 1997)}
\end{center}
\bigskip
\bigskip
\bigskip
\bigskip
\bigskip
\bigskip
\bigskip
\bigskip
\bigskip
\bigskip
\bigskip
\bigskip
\bigskip
\bigskip
\bigskip
\bigskip
\bigskip
\newpage
\setcounter{page}{1}
\pagestyle{plain}
\noindent
\hspace*{0.3cm}
The phenomenon of flavor mixing, which is intrinsically linked to
$CP$--violation, is an important ingredient of the Standard Model of Basic
Interactions. Yet unlike other features of the Standard Model, e.\ g.\ the
mixing of the neutral electroweak gauge bosons, it is a phenomenon which can
merely be described. A deeper understanding is still lacking, but most
theoreticians would agree that it is directly linked to the mass spectrum of
the quarks -- the possible mixing of lepton flavors will not be discussed
here. Furthermore there is a general consensus that a deeper dynamical
understanding would require to go beyond the physics of the Standard Model.
In this talk I shall not go thus far. Instead I shall demonstrate that the
observed properties of the flavor mixing, combined with our knowledge about
the quark mass spectrum, suggest specific symmetry properties which allow
to fix the flavor mixing parameters with high precision, thus predicting the
outcome of the experiments which will soon be performed at the $B$--meson
factories.\\
\\
\hspace*{0.3cm} In the standard electroweak theory, the phenomenon of flavor
mixing of the quarks is described by a $3\times 3$ unitary matrix, the
Cabibbo-Kobayashi-Maskawa (CKM) matrix$^{1,2}$. This
matrix can be expressed in terms of four parameters, which are usually 
taken as three rotation angles and one phase. A number of
different parametrizations have been proposed in the
literature$^{2,3,4,5}$. Of course, adopting a particular parametrization
of flavor mixing is arbitrary and not directly a physics
issue. Nevertheless it is quite likely that the actual values of 
flavor mixing parameters (including the strength of $CP$ violation),
once they are known with high precision, will give interesting information 
about the physics beyond the standard model. Probably at this point it 
will turn out that a particular description of the CKM matrix is more
useful and transparent than the others. For this reason, let me first 
analyze all possible parametrizations and point out their
respective advantages and disadvantages.\\
\\
\hspace*{0.3cm} In the standard model the quark flavor mixing arises once
the up- and
down-type mass matrices are diagonalized. The generation of quark masses 
is intimately related to the phenomenon of flavor mixing. In
particular, the flavor mixing parameters do depend on the
elements of quark mass matrices. A particular structure of the
underlying mass matrices calls for a particular choice of the
parametrization of the flavor mixing matrix. For example, in
it was noticed$^{6}$ that a rather special form of
the flavor mixing matrix results, if one starts from Hermitian mass
matrices in which the (1,3) and (3,1) elements vanish. This has been
subsequently observed again in a number of papers$^{7}$. Recently we have
studied the exact form of such a
description from a general point of view and pointed out many
advantages of this type of representation in the discussion of flavor
mixing and $CP$-violating phenomena$^{5}$, which will be discussed
later.\\
\\
\hspace*{0.3cm} In the standard model the weak charged currents are given by 
\begin{equation}
\overline{(u, ~ c, ~ t)}^{~}_L \left ( \matrix{
V_{ud}  & V_{us}        & V_{ub} \cr
V_{cd}  & V_{cs}        & V_{cb} \cr 
V_{td}  & V_{ts}        & V_{tb} \cr} \right ) 
\left ( \matrix{
d \cr s \cr b \cr} \right  )_L \; ,
\end{equation}
where $u$, $c$, ..., $b$ are the quark mass eigenstates, $L$ denotes
the left-handed fields, and $V_{ij}$ are elements of the CKM matrix
$V$. In general $V_{ij}$ are complex numbers, but their absolute
values are measurable quantities. For example, $|V_{cb}|$ primarily
determines the lifetime of $B$ mesons. The phases of $V_{ij}$,
however, are not physical, like the phases of quark fields. A phase
transformation of the $u$ quark ($u \rightarrow u ~ e^{{\rm
i}\alpha}$), for example, leaves the quark mass term invariant but
changes the elements in the first row of $V$ (i.e., $V_{uj} \rightarrow 
V_{uj} ~ e^{-{\rm i}\alpha}$). Only a common phase transformation of all 
quark fields leaves all elements of $V$ invariant, thus there is a
five-fold freedom to adjust the phases of $V_{ij}$.\\
\\
\hspace*{0.3cm}
In general the unitary matrix $V$ depends on nine parameters.
Note that in the absence of complex phases $V$ would consist of only three 
independent parameters, corresponding to three (Euler) rotation
angles. Hence one can describe the complex matrix $V$ by three
angles and six phases. Due to the freedom in redefining the quark
field phases, five of the six phases in $V$ can be absorbed, and we arrive
at the well-known result that the CKM matrix $V$ can be parametrized
in terms of three rotation angles and one $CP$-violating phase. The
question about how many different ways to describe $V$ may exist was
raised some time ago$^{8}$. Recently the problem was
reconsidered and brought in connection with the mass hierarchy$^{5}$.\\
\\
\hspace*{0.3cm} In our view the best possibility to describe the flavor
mixing in the
standard model is to adopt the parametrization discussed in
ref.\ (5). This parametrization has a number of significant
advantages.
In the following part I shall show that this parametrization
follows
automatically if we impose the constraints from the chiral symmetries and the
hierarchical structure of the mass eigenvalues$^{9, 10, 11}$. We take the
point of view
that the quark mass eigenvalues are dynamical entities, and one could change
their values in order to study certain
symmetry limits, as it is done in QCD. In the standard electroweak model, in
which the quark mass matrices are given by the coupling of a scalar field to
various quark fields, this can certainly be done by adjusting the related
coupling constants. Whether it is possible in reality is an open question. It
is well--known that the quark mass matrices can always be made hermitian
by a suitable transformation of the right--handed fields. Without loss of
generality, we shall suppose in this paper that the quark mass matrices are
hermitian. In the limit where the masses of the $u$ and $d$ quarks are set to
zero, the quark mass matrix $\tilde{M}$ (for both charge $+2/3$ and
charge $-1/3$ sectors) can be arranged such that its elements 
$\tilde{M}_{i1}$ and $\tilde{M}_{1i}$ ($i=1,2,3$) are all zero$^{9, 10}$.
Thus the quark mass matrices have the form
\begin{equation}
\tilde{M} \; =\; \left ( \matrix{
0       & 0     & 0 \cr
0       & \tilde{C}     & \tilde{B} \cr
0       & \tilde{B}^*   & \tilde{A} \cr} \right ) \; .
\end{equation}
\hspace*{0.3cm} The observed mass hierarchy is incorporated into this
structure by denoting the entry which is of the order of the $t$-quark or 
$b$-quark mass by $\tilde{A}$, with $\tilde{A}\gg \tilde{C},
|\tilde{B}|$. It can easily be seen (see, e.g., ref.\ (12) that
the complex phases in the mass matrices (1) can be
rotated away by subjecting both $\tilde{M}_{\rm u}$ and $\tilde{M}_{\rm d}$ to the
same unitary transformation. Thus we shall take $\tilde{B}$ to be
real for both up- and down-quark sectors. As expected, $CP$ violation
cannot arise at this stage. The diagonalization of the mass matrices
leads to a mixing between the second and third families, described by an
angle $\tilde{\theta}$. The flavor mixing matrix is
then given by
\begin{equation}
\tilde{V} \; =\; \left ( \matrix{
1       & 0     & 0 \cr
0       & \tilde{c}     & \tilde{s} \cr
0       & -\tilde{s}    & \tilde{c} \cr } \right ) \; ,
\end{equation}
where $\tilde{s} \equiv \sin \tilde{\theta}$ and $\tilde{c} \equiv
\cos \tilde{\theta}$. In view of the fact that the limit $m_u = m_d
=0$ is not far from reality, the angle $\tilde{\theta}$ is essentially 
given by the observed value of $|V_{cb}|$
($=0.039 \pm 0.002$)$^{13, 14}$;
i.e., $\tilde{\theta} = 2.24^{\circ} \pm 0.12^{\circ}$.\\ 
\\
\hspace*{0.3cm} At the next and final stage of the chiral evolution of the
mass matrices, the masses of the $u$ and $d$ quarks are introduced.
The Hermitian mass matrices have in general the
form:
\begin{equation}
M \; =\; \left ( \matrix{
E       & D     & F \cr
D^*     & C     & B \cr
F^*     & B^*   & A \cr } \right ) \; 
\end{equation}
with $A\gg C, |B| \gg E, |D|, |F|$. By a common unitary transformation of 
the up- and down-type quark fields, one can always arrange the mass
matrices $M_{\rm u}$ and $M_{\rm d}$ in such a way that $F_{\rm u} =
F_{\rm d} =0$; i.e.,
\begin{equation}
M \; =\; \left ( \matrix{
E       & D     & 0 \cr
D^*     & C     & B \cr
0       & B^*   & A \cr } \right ) \; .
\end{equation}
\hspace*{0.3cm} This can easily be seen as follows. If phases are neglected,
the two symmetric mass matrices $M_{\rm u}$ and $M_{\rm d}$ can be transformed 
by an orthogonal transformation matrix $O$, which can be described by
three angles such that they assume the form (5). The condition
$F_{\rm u} =F_{\rm d} =0$ gives two constraints for the three angles of 
$O$. If complex phases are allowed in $M_{\rm u}$ and $M_{\rm d}$, the 
condition $F_{\rm u} =F_{\rm u}^* = F_{\rm d} =F_{\rm d}^* =0$ imposes
four constraints, which can also be fulfilled, if $M_{\rm u}$ and
$M_{\rm d}$ are subjected to a common unitary transformation matrix $U$. The
latter depends on nine parameters. Three of them are not suitable for
our purpose, since they are just diagonal phases; but the remaining
six can be chosen such that the vanishing of $F_{\rm u}$ and $F_{\rm
d}$ results.\\ 
\\
\hspace*{0.3cm} The basis in which the mass matrices take the form (5) is a
basis in the space of quark flavors, which in our view is of special
interest. It is a basis in which the mass matrices exhibit two
texture zeros, for both up- and down-type quark sectors. 
These, however, do not imply special relations among
mass eigenvalues and flavor mixing parameters (as pointed out
above). In this basis the mixing is of the ``nearest neighbour'' form, 
since the (1,3) and (3,1) elements of $M_{\rm u}$ and $M_{\rm d}$
vanish; no direct mixing between the heavy $t$ (or $b$) quark and the
light $u$ (or $d$) quark is present (see also ref.\ (15).
In certain models (see, e.g., ref.\ (16),
this basis is indeed of particular interest, but we shall proceed without 
relying on a special texture models for the mass matrices.\\
\\
\hspace*{0.3cm} A mass matrix of the type (5) can in the absence of complex
phases be diagonalized by a rotation matrix, described by two angles only.
At first the off-diagonal element 
$B$ is rotated away by a rotation between the second and third 
families (angle $\theta_{23}$); at the second step the element $D$ is rotated away by a
transformation of the first and second families (angle $\theta_{12}$). No rotation between
the first and third families is required. 
The rotation matrix for this sequence takes the form
\begin{eqnarray}
R \; =\; R_{12} R_{23} & = & \left ( \matrix{
c_{12}  & s_{12}        & 0 \cr
-s_{12} & c_{12}        & 0 \cr
0       & 0     & 1 \cr } \right )  \left ( \matrix{
1       & 0     & 0 \cr
0       & c_{23}        & s_{23} \cr
0       & -s_{23}       & c_{23} \cr } \right ) \; 
\nonumber \\ \nonumber \\
& = & \left ( \matrix{
c_{12}  & s_{12} c_{23} & s_{12} s_{23} \cr 
-s_{12} & c_{12} c_{23} & c_{12} s_{23} \cr
0       & -s_{23}       & c_{23} \cr } \right ) \; ,
\end{eqnarray}
where $c_{12} \equiv \cos \theta_{12}$, $s_{12} \equiv \sin
\theta_{12}$, etc.
The flavor mixing matrix $V$ is the product of two such matrices, one
describing the rotation among the up-type quarks, and the other describing
the rotation among the down-type quarks:
\begin{equation}
V \; =\; R^{\rm u}_{12} R^{\rm u}_{23} ( R^{\rm d}_{23} )^{-1} ( R^{\rm d}_{12} )^{-1} \; .
\end{equation}
\hspace*{0.3cm} The product $R^{\rm u}_{23} (R^{\rm d}_{23} )^{-1}$ can be
written as a rotation matrix described by a single angle $\theta$. In the
limit $m_u = m_d =0$, this is just the angle $\tilde{\theta}$ encountered
in eq. (6). The angle which describes the $R^{\rm u}_{12}$ rotation shall
be denoted by $\theta_{\rm u}$; the corresponding angle for the
$R^{\rm d}_{12}$ rotation by $\theta_{\rm d}$. Thus in the absence of
$CP$-violating phases the flavor mixing matrix takes the following 
specific form:
\begin{eqnarray}
V & = & \left ( \matrix{
c_{\rm u}       & s_{\rm u}     & 0 \cr
-s_{\rm u}      & c_{\rm u}     & 0 \cr
0       & 0     & 1 \cr } \right )  \left ( \matrix{
1       & 0     & 0 \cr
0       & c     & s \cr
0       & -s    & c \cr } \right )  \left ( \matrix{
c_{\rm d}       & -s_{\rm d}    & 0 \cr
s_{\rm d}       & c_{\rm d}     & 0 \cr
0       & 0     & 1 \cr } \right )  \nonumber \\ \nonumber \\
& = & \left ( \matrix{
s_{\rm u} s_{\rm d} c + c_{\rm u} c_{\rm d}     & 
s_{\rm u} c_{\rm d} c - c_{\rm u} s_{\rm d}     & s_{\rm u} s \cr
c_{\rm u} s_{\rm d} c - s_{\rm u} c_{\rm d}     & 
c_{\rm u} c_{\rm d} c + s_{\rm u} s_{\rm d}     & c_{\rm u} s \cr
-s_{\rm d} s    & -c_{\rm d} s  & c \cr } \right ) \; ,
\end{eqnarray}
where $c_{\rm u} \equiv \cos\theta_{\rm u}$, $s_{\rm u} \equiv
\sin\theta_{\rm u}$, etc.\\
\\
\hspace*{0.3cm} We proceed by including the phase parameters of the quark
mass matrices in eq.\ (5). It can easily be seen that, 
by suitable rephasing of the quark fields,
the flavor mixing matrix can finally be written in terms of only a
single phase $\varphi$ as follows:
\begin{eqnarray}
V & = & \left ( \matrix{
c_{\rm u}       & s_{\rm u}     & 0 \cr
-s_{\rm u}      & c_{\rm u}     & 0 \cr
0       & 0     & 1 \cr } \right )  \left ( \matrix{
e^{-{\rm i}\varphi}     & 0     & 0 \cr
0       & c     & s \cr
0       & -s    & c \cr } \right )  \left ( \matrix{
c_{\rm d}       & -s_{\rm d}    & 0 \cr
s_{\rm d}       & c_{\rm d}     & 0 \cr
0       & 0     & 1 \cr } \right )  \nonumber \\ \nonumber \\
& = & \left ( \matrix{
s_{\rm u} s_{\rm d} c + c_{\rm u} c_{\rm d} e^{-{\rm i}\varphi} &
s_{\rm u} c_{\rm d} c - c_{\rm u} s_{\rm d} e^{-{\rm i}\varphi} &
s_{\rm u} s \cr
c_{\rm u} s_{\rm d} c - s_{\rm u} c_{\rm d} e^{-{\rm i}\varphi} &
c_{\rm u} c_{\rm d} c + s_{\rm u} s_{\rm d} e^{-{\rm i}\varphi}   &
c_{\rm u} s \cr
- s_{\rm d} s   & - c_{\rm d} s & c \cr } \right ) \; .
\end{eqnarray}
\hspace*{0.3cm} Note that the three angles $\theta_{\rm u}$, $\theta_{\rm d}$
and $\theta$ in eq. (12) can all be arranged to lie in the first quadrant
through a suitable redefinition of quark field phases. Consequently
all $s_{\rm u}$, $s_{\rm d}$, $s$ and $c_{\rm u}$, $c_{\rm d}$, $c$
are positive. The phase $\varphi$ can in general take values from 0
to $2\pi$; and $CP$ violation is present in the weak interactions
if $\varphi \neq 0, \pi$ and $2\pi$.\\
\\
\hspace*{0.3cm} This particular representation of the flavor mixing matrix is
the main result of this paper. In comparison with all other parametrizations
discussed previously$^{2, 3}$, it
has a number of interesting features which in our view make it very
attractive and provide strong arguments for its use in future
discussions of flavor mixing phenomena, in particular, those in
$B$-meson physics (see also refs.\ (17, 18)). We shall discuss them below.\\
\\
a) The flavor mixing matrix $V$ in eq. (12) follows directly from the
chiral expansion of the mass
matrices. Thus it naturally takes into account the hierarchical structure of the 
quark mass spectrum.\\
\\
b) The complex phase $\varphi $ describing $CP$ violation appears only in the
(1,1), (1,2), (2,1) and (2,2) elements of $V$, i.e., 
in the elements involving only the quarks of the first and second
families. This is a natural description of $CP$ violation since in our 
hierarchical approach $CP$ violation is not directly linked to the third family, but
rather to the first and second ones, and in particular to the mass terms of the
$u$ and $d$ quarks.\\ 
\\
\hspace*{0.3cm} It is instructive to consider the special case
$s_{\rm u} = s_{\rm d} = s = 0$. Then the flavor mixing matrix $V$ takes the
form
\begin{equation}
V \; = \; \left ( \matrix{
e^{-{\rm i}\varphi}     & 0     & 0 \cr
0       & 1     & 0 \cr
0       & 0     & 1 \cr} \right ) \; .
\end{equation}
\hspace*{0.3cm} This matrix describes a phase change in the weak transition
between $u$ and $d$, while no phase change is present in the
transitions between $c$ and $s$ as well as $t$ and $b$.
Of course, this effect can be absorbed in a phase change of the $u$-
and $d$-quark fields, and no $CP$ violation is present. Once the
angles $\theta_{\rm u}$, $\theta_{\rm d}$ and $\theta$ are introduced, 
however, $CP$ violation arises. It is due to a phase change in the weak
transition between $u^{\prime}$ and $d^{\prime}$, where $u^{\prime}$
and $d^{\prime}$ are the rotated quark fields, obtained by applying
the corresponding rotation matrices given in eq. (9) to the 
quark mass eigenstates ($u^{\prime}$: mainly $u$, small admixture of
$c$; $d^{\prime}$: mainly $d$, small admixture of $s$).\\
\\
\hspace*{0.3cm} Since the mixing matrix elements involving the $t$ or $b$
quarks
are real in the representation (9), one can find that the phase parameter of
$B^0_q$-$\bar{B}^0_q$ mixing ($q=d$ or $s$), dominated by the
box-diagram contributions in the standard model$^{19}$, is essentially unity:
\begin{equation}
\left ( \frac{q}{p} \right )_{B_q} = \;
\frac{V^*_{tb}V_{tq}}{V_{tb}V^*_{tq}} \; = \; 1 \; .
\end{equation}
\hspace*{0.3cm} In most other parametrizations of the flavor mixing
matrix, however, the two rephasing-variant quantities 
$(q/p)^{~}_{B_d}$ and $(q/p)^{~}_{B_s}$ take different (maybe complex)
values.\\
\\
c) The dynamics of flavor mixing can easily be interpreted by
considering certain limiting cases in eq. (9). In the limit $\theta
\rightarrow 0$ (i.e., $s \rightarrow 0$ and $c\rightarrow 1$), the
flavor mixing is, of course, just a mixing between the first and
second families, described by only one mixing angle (the Cabibbo angle 
$\theta_{\rm C}$$^{1}$).\\  
It is a special and essential feature of the representation (9) that the Cabibbo
angle is {\it not} a basic angle, used in the parametrization. 
The matrix element $V_{us}$ (or $V_{cd}$) is
indeed a superposition of two terms including a phase. This feature
arises naturally in our hierarchical approach, but it is not new. In
many models of specific textures of mass matrices, it is indeed the
case that the Cabibbo-type transition $V_{us}$ (or $V_{cd}$) 
is a superposition of several
terms. At first, it was obtained by one of the authors 
in the discussion of the two-family mixing$^{20}$.\\
\\
\hspace*{0.3cm} In the limit $\theta =0$ considered here, one has
$|V_{us}| = |V_{cd}| = \sin\theta_{\rm C} \equiv s^{~}_{\rm C}$ and
\begin{equation}
s^{~}_{\rm C} \; =\; \left | s_{\rm u} c_{\rm d} ~ - ~ c_{\rm u} s_{\rm d}
e^{-{\rm i}\varphi} \right | \; .
\end{equation}
\hspace*{0.3cm} This relation describes a triangle in the complex plane, as
illustrated in Fig. 1, which we shall denote as the ``LQ-- triangle''
(``light quark triangle''). 
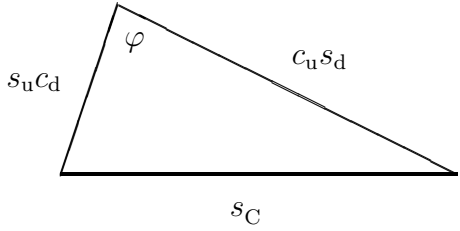
\begin{figure}[t]
\begin{picture}(400,160)(-90,210)
\put(80,300){\line(1,0){150}}
\put(80,300.5){\line(1,0){150}}
\put(150,285.5){\makebox(0,0){$s^{~}_{\rm C}$}}
\put(80,300){\line(1,3){21.5}}
\put(80,300.5){\line(1,3){21.5}}
\put(80,299.5){\line(1,3){21.5}}
\put(70,335){\makebox(0,0){$s_{\rm u} c_{\rm d}$}}
\put(230,300){\line(-2,1){128}}
\put(230,300.5){\line(-2,1){128}}
\put(178,343.5){\makebox(0,0){$c_{\rm u} s_{\rm d}$}}

\put(108,350){\makebox(0,0){$\varphi$}}
\end{picture}
\vspace{-2.5cm}
\caption{The LQ--triangle in the complex plane.}
\end{figure}
\hspace*{0.3cm} This triangle is a feature of
the mixing of the first two families (see also ref.\ (20)). Explicitly one has
(for $s=0$):
\begin{equation}
\tan\theta_{\rm C} \; =\; \sqrt{\frac{\tan^2\theta_{\rm u} +
\tan^2\theta_{\rm d} - 2 \tan\theta_{\rm u} \tan\theta_{\rm d}
\cos\varphi}
{1 + \tan^2\theta_{\rm u} \tan^2\theta_{\rm d} + 2 \tan\theta_{\rm u}
\tan\theta_{\rm d} \cos\varphi}} \; .
\end{equation}
\hspace*{0.3cm} Certainly the flavor mixing matrix $V$ cannot accommodate
$CP$ violation in this
limit. However, the existence of $\varphi$ seems necessary in order
to make eq. (16) compatible with current data, as one can see below.\\
\\
d) The three mixing angles $\theta$, $\theta_{\rm u}$ and 
$\theta_{\rm d}$ have a precise physical meaning. The angle $\theta$
describes the mixing between the second and third families, which is
generated by the off-diagonal terms $B_{\rm u}$ and $B_{\rm d}$ in the 
up and down mass matrices of eq. (9). 
We shall refer to this mixing involving $t$ and $b$ as the ``heavy
quark mixing''.
The angle $\theta_{\rm u}$,
however, primarily describes the $u$-$c$ mixing, corresponding to the
$D_{\rm u}$ term in $M_{\rm u}$. We shall denote this as the ``u-channel
mixing''. The angle $\theta_{\rm d}$ primarily describes 
the $d$-$s$ mixing, corresponding to the $D_{\rm d}$ term in $M_{\rm
d}$; this will be denoted as the ``d-channel mixing''. 
Thus there exists an asymmetry between the mixing of the first and
second families and that of the second and third families,
which in our view reflects interesting details of the underlying dynamics of
flavor mixing. 
The heavy quark mixing is a combined effect, involving both charge
$+2/3$ and charge $-1/3$ quarks, while the u- or d-channel mixing
(described by the angle $\theta_{\rm u}$ or $\theta_{\rm d}$) proceeds 
solely in the charge $+2/3$ or charge $-1/3$ sector. Therefore an
experimental determination of these two angles would allow to draw
interesting conclusions about the amount and perhaps the underlying
pattern of the u- or d-channel mixing.\\
\\
e) The three angles $\theta$, $\theta_{\rm u}$ and $\theta_{\rm d}$
are related in a very simple way to observable quantities of $B$-meson 
physics.\\ 
For example, $\theta$ is related to 
the rate of the semileptonic decay $B\rightarrow D^*l\nu^{~}_l$; 
$\theta_{\rm u}$ is associated with the ratio of the decay rate of
$B\rightarrow (\pi, \rho) l \nu^{~}_l$ to that of $B\rightarrow 
D^* l\nu^{~}_l$; and $\theta_{\rm d}$ can be determined from the ratio of
the mass difference between two $B_d$ mass eigenstates to that between
two $B_s$ mass eigenstates. From eq. (9) we find the following exact
relations:
\begin{equation}
\sin \theta \; = \; |V_{cb}| \sqrt{ 1 + \left |\frac{V_{ub}}{V_{cb}}
\right |^2} \; ,
\end{equation}
and
\begin{eqnarray}
\tan\theta_{\rm u} & = & \left | \frac{V_{ub}}{V_{cb}} \right | \; ,
\nonumber \\
\tan\theta_{\rm d} & = & \left | \frac{V_{td}}{V_{ts}} \right | \; .
\end{eqnarray}
These simple results make the parametrization (9) uniquely favorable 
for the study of $B$-meson physics.\\
\\
\hspace*{0.3cm} By use of current data on $|V_{ub}|$ and $|V_{cb}|$, i.e.,
$|V_{cb}| = 0.039 \pm 0.002$$^{13, 14}$ and $|V_{ub}/V_{cb}|
= 0.08 \pm 0.02$$^{14}$, we obtain $\theta_{\rm u} = 4.57^{\circ} \pm
1.14^{\circ}$ and $\theta = 2.25^{\circ} \pm 0.12^{\circ}$. Taking
$|V_{td}| = (8.6 \pm 2.1) \times 10^{-3}$$^{14}$,
which was obtained from the analysis of current data on
$B^0_d$-$\bar{B}^0_d$ mixing,
we get $|V_{td}/V_{ts}| = 0.22 \pm 0.07$, i.e., $\theta_{\rm d} = 12.7^{\circ} 
\pm 3.8^{\circ}$.
\hspace*{0.3cm}Both the heavy quark mixing angle $\theta$ and the u-channel
mixing angle $\theta_{\rm u}$ are relatively small. The smallness of $\theta$ 
implies that Eqs. (11) and (12) are valid to a high degree of
precision (of order $1-c \approx 0.001$).\\
\\
f) According to eq. (12), as well as eq. (11), the phase $\varphi$ is
a phase difference between the contributions to $V_{us}$ (or $V_{cd}$) 
from the u-channel mixing and the d-channel mixing. Therefore
$\varphi$ is given by the relative phase of $D_{\rm d}$ and $D_{\rm
u}$ in the quark mass matrices (4), if the phases of $B_{\rm u}$ and
$B_{\rm d}$ are absent or negligible.\\ 
\\
\hspace*{0.3cm} The phase $\varphi$ is not likely to be $0^{\circ}$ or
$180^{\circ}$, according
to the experimental values given above, even though the measurement of 
$CP$ violation in $K^0$-$\bar{K}^0$ mixing$^{19}$ is not taken
into account. For $\varphi =0^{\circ}$, one
finds $\tan\theta_{\rm C} = 0.14 \pm 0.08$; and for $\varphi =
180^{\circ}$, one gets $\tan\theta_{\rm C} = 0.30 \pm 0.08$. Both
cases are barely consistent with the value of $\tan\theta_{\rm
C}$ obtained from experiments ($\tan\theta_{\rm C} \approx
|V_{us}/V_{ud}| \approx 0.226$).\\ 
\\
g) The $CP$-violating phase $\varphi$ in the flavor mixing matrix $V$ can be
determined from $|V_{us}|$ ($= 0.2205 \pm 0.0018$)$^{19}$
through the following formula, obtained easily from eq. (8):
\begin{equation}
\varphi \; =\; \arccos \left ( \frac{s^2_{\rm u} c^2_{\rm d} c^2 +
c^2_{\rm u} s^2_{\rm d} - |V_{us}|^2}{2 s_{\rm u} c_{\rm u} s_{\rm d}
c_{\rm d} c} \right ) \; .
\end{equation}
\\
\hspace*{0.3cm} The two--fold ambiguity associated with the value of
$\varphi$, coming
from $\cos\varphi = \cos (2\pi - \varphi)$, is removed if one
takes $\sin\varphi >0$ into account (this is required by current data on
$CP$ violation in $K^0$-$\bar{K}^0$ mixing (i.e., $\epsilon^{~}_K$).
More precise measurements of the angles $\theta_{\rm u}$ and
$\theta_{\rm d}$ in the forthcoming experiments of $B$ physics will
reduce the uncertainty of $\varphi$ to be determined from eq.
(19). This approach is of course complementary to the direct determination of
$\varphi$ from $CP$ asymmetries in some weak $B$-meson decays into hadronic
$CP$ eigenstates$^{21}$. 
\newpage
Considering the presently known phenomenological constraints (see e.g.
Ref.$^{22}$) the value of $\varphi$ is most likely in the range
$40^{\circ}$ to $120^{\circ}$; the central value is
$\varphi \approx 81^{\circ}$. 
Note that $\varphi$ is essentially independent of the angle $\theta$,
due to the tiny observed value of the latter. 
Once $\tan\theta_{\rm d}$ is precisely measured, one shall be able to fix the
magnitude of $\varphi$ to a satisfactory degree of accuracy.\\
h) It is well--known that $CP$ violation in the flavor mixing matrix $V$ can
be rephasing--invariantly described by a universal quantity ${\cal J}$$^{23}$:
\begin{equation}
{\rm Im} \left( V_{il} V_{jm} V^*_{im} V^*_{jl} \right) = {\cal J}
\sum\limits^{3}_{k,n=1} \left( \epsilon_{ijk}\epsilon_{lmn} \right] \, .
\end{equation}
In the parametrisation (9), ${\cal J}$ reads
\begin{equation}
{\cal J} = s_uc_us_dc_ds^2 c sin \varphi
\end{equation}
Obviously $\varphi = 90^{\circ }$ leads to the maximal value of ${\cal J}$.
Indeed $\varphi =90^{\circ}$, a particularly interesting case for $CP$ 
violation, is quite consistent with
current data.
In this case the mixing term
$D_{\rm d}$ in eq. (5) can be taken to be real, and the term $D_{\rm 
u}$ to be imaginary, if ${\rm Im}(B_{\rm u}) = {\rm Im} (B_{\rm d})
=0$ is assumed. 
Since in our description of the flavor mixing the
complex phase $\varphi$ is related in a simple way to the phases of
the quark mass terms, the case $\varphi = 90^{\circ}$ is especially
interesting. It can hardly be an accident, and this case should be
studied further. The possibility that the phase $\varphi$ describing
$CP$ violation in the standard model is given by the algebraic number
$\pi/2$ should be taken seriously. It may provide a useful clue
towards a deeper understanding of the origin of $CP$ violation
and of the dynamical origin of the fermion masses.\\
\\
\hspace*{0.3cm} In ref.\ (5) the case $\varphi =90^{\circ}$ has been
denoted as ``maximal'' $CP$ violation. It implies in our framework 
that in the complex
plane the u-channel and d-channel mixings are perpendicular to each
other. In this special case (as well as $\theta\rightarrow 0$), we have 
\begin{equation}
\tan^2\theta_{\rm C} \; =\; \frac{\tan^2\theta_{\rm u} ~ + ~
\tan^2\theta_{\rm d}}{1 ~ + ~ \tan^2\theta_{\rm u} \tan^2\theta_{\rm
d}} \; .
\end{equation}
To a good approximation (with the relative error $\sim 2\%$), 
one finds $s^2_{\rm C} \approx s^2_{\rm u} + s^2_{\rm d}$.\\ 
\\
i) At future $B$-meson factories, the study of $CP$ violation will
concentrate on measurements of the unitarity triangle.\\ 
The unitzarity triangle (a) and its rescaled counterpart (b) in the complex
plane.
\begin{equation}
S_u ~ + ~ S_c ~ + ~ S_t \; = \; 0 \; ,
\end{equation}
where $S_i \equiv V_{id} V^*_{ib}$ in the complex
plane (see Fig. 2(a)). The inner angles of this triangle.
are denoted as$^{19}$.
\begin{eqnarray}
\alpha & \equiv & \arg (- S_t S^*_u ) \; , \nonumber \\
\beta  & \equiv & \arg (- S_c S^*_t ) \; , \nonumber \\
\gamma & \equiv & \arg (- S_u S^*_c ) \; .
\end{eqnarray}

In terms of the parameters
$\theta$, $\theta_{\rm u}$, $\theta_{\rm d}$ and $\varphi$, we obtain
\begin{eqnarray}
\sin (2\alpha) & = & \frac{2 c_{\rm u} c_{\rm d} \sin\varphi \left
( s_{\rm u} s_{\rm d} c + c_{\rm u} c_{\rm d} \cos\varphi \right )}{s^2_{\rm
u} s^2_{\rm d} c^2 + c^2_{\rm u} c^2_{\rm d} + 2 s_{\rm u} c_{\rm u} s_{\rm d} c_{\rm d} c
\cos\varphi} \; , \nonumber \\ \nonumber \\
\sin (2\beta) & = & \frac{2 s_{\rm u} c_{\rm d} \sin\varphi \left
( c_{\rm u} s_{\rm d} c - s_{\rm u} c_{\rm d} \cos\varphi \right )}{c^2_{\rm
u} s^2_{\rm d} c^2 + s^2_{\rm u} c^2_{\rm d} - 2 s_{\rm u} c_{\rm u} s_{\rm d} c_{\rm d} c
\cos\varphi} \; .
\end{eqnarray}
To an excellent degree of accuracy, one finds $\alpha \approx
\varphi$. In order to illustrate how accurate this relation is, let us
input the central values of $\theta$, $\theta_{\rm u}$ and $\theta_{\rm 
d}$ (i.e., $\theta = 2.25^{\circ}$, $\theta_{\rm u} = 4.57^{\circ}$
and $\theta_{\rm d} = 12.7^{\circ}$) to eq. (22). Then one arrives at
$\varphi - \alpha \approx 1^{\circ}$ as well as $\sin (2\alpha)
\approx 0.34$ and $\sin (2\beta) \approx 0.65$. 
It is expected that $\sin (2\alpha)$ and $\sin (2\beta)$
will be directly measured from the $CP$ asymmetries in 
$B_d \rightarrow \pi^+\pi^-$ and $B_d \rightarrow J /\psi K_S$ modes
at a $B$-meson factory.\\
\\
\hspace*{0.3cm} Note that the three sides of the unitarity triangle 
(21) can be rescaled by $|V_{cb}|$. In a very good approximation
(with the relative error $\sim 2\%$), one arrives at
\begin{equation}
|S_u| ~ : ~ |S_c| ~ : ~ |S_t| \; \approx \; s_{\rm u} c_{\rm d} ~ : ~ 
s^{~}_{\rm C} ~ : ~ s_{\rm d} \; .
\end{equation}
Equivalently, one can obtain
\begin{equation}
s_{\alpha} ~ : ~ s^{~}_{\beta} ~ : ~ s_{\gamma} \; \approx \; s^{~}_{\rm C} 
~ : ~ s_{\rm u} c_{\rm d} ~ : ~ s_{\rm d} \; ,
\end{equation}
where $s_{\alpha} \equiv \sin\alpha$, etc.
Comparing the
unitarity triangle with the LQ--triangle in Fig. 1, we find that they are 
indeed congruent with each other to a high degree of accuracy.
The congruent relation between these two triangles is particularly
interesting, since the LQ--triangle is essentially a feature of the physics
of the first two quark families, while the unitarity triangle is
linked to all three families. In this connection it is of special
interest to note that in models which specify the textures of the mass 
matrices the Cabibbo triangle and hence three inner angles of the unitarity
triangle can be fixed by the spectrum of the light quark masses and
the $CP$-violating phase $\varphi$.

j) It is worth pointing out that the u-channel and d-channel mixing
angles are related to the so-called Wolfenstein parameters$^{23}$ in a simple
way:
\begin{eqnarray}
\tan\theta_{\rm u} & = & \left | \frac{V_{ub}}{V_{cb}} \right | 
\; \approx \; \lambda \sqrt{\rho^2 + \eta^2} \; , \; \nonumber \\
\tan\theta_{\rm d} & = & \left | \frac{V_{td}}{V_{ts}} \right |
\; \approx \; \lambda \sqrt{ (1-\rho)^2 + \eta^2} \; ,
\end{eqnarray}
where $\lambda \approx s^{~}_{\rm C}$ measures the magnitude of $V_{us}$.
Note that the $CP$-violating parameter $\eta$ is linked to $\varphi$
through
\begin{equation}
\sin\varphi \; \approx \; \frac{\eta}{\sqrt{\rho^2 + \eta^2}
\sqrt{(1-\rho)^2 + \eta^2}} \; 
\end{equation}
in the lowest-order approximation. Then $\varphi =90^{\circ}$ implies
$\eta^2 \approx \rho ( 1- \rho)$, on the condition $0 < \rho < 1$. In
this interesting case, of course, the flavor mixing matrix can 
fully be described in terms of only three independent parameters.

k) Compared with the standard parametrization of the flavor mixing
matrix $V$ the parametrization (9) has an additional
advantage: the renormalization-group evolution of $V$, from the weak
scale to an arbitrary high energy scale, is 
to a very good approximation associated only with the angle $\theta$. This can
easily be seen if one keeps the $t$ and $b$ Yukawa couplings only  
and neglects possible threshold effect in the one-loop
renormalization-group equations of the Yukawa matrices$^{24}$.
Thus the parameters $\theta_{\rm u}$, $\theta_{\rm d}$ and $\varphi$
are essentially independent of the energy scale, while $\theta$ does
depend on it and will change if the underlying scale is shifted, say
from the weak scale ($\sim 10^2$ GeV) to the grand unified theory
scale (of order $ 10^{16}$ GeV). In short, the heavy quark mixing is
subject to renormalization-group effects; but the u- and d-channel
mixings are not, likewise the phase $\varphi$ describing $CP$
violation and the LQ--triangle as a whole.\\
\\
\hspace*{0.3cm} We have presented a new description of the flavor mixing 
phenomenon, which is based on the phenomenological fact that the quark 
mass spectrum exhibits a clear hierarchy pattern. This leads uniquely
to the interpretation of the flavor mixing in terms of a heavy quark
mixing, followed by the u-channel and d-channel mixings. The complex
phase $\varphi$, describing the relative orientation of the u-channel
mixing and the d-channel mixing in the complex plane, signifies
$CP$ violation, which is a phenomenon primarily linked to the physics
of the first two families. The Cabibbo angle is not a basic mixing
parameter, but given by a superposition of two terms involving the
complex phase $\varphi$. The experimental data suggest that the phase
$\varphi$, which is directly linked to the phases of the quark mass
terms, is close to $90^{\circ}$. This opens the possibility to
interpret $CP$ violation as a maximal effect, in a similar way as
parity violation.\\ 
\\
\hspace*{0.3cm} Our description of flavor mixing has many clear advantages
compared with other descriptions. We propose that it should be used in the
future description of flavor mixing and $CP$ violation, in particular, 
for the studies of quark mass matrices and $B$-meson physics.\\
\\
\hspace*{0.3cm} The description of the flavor mixing phenomenon given above
is of special
interest if for the $U$ and $D$ channel mixing the quark mass textures
discussed first in$^{20}$ are applied (see also$^{5}$).
In that case one finds$^{25}$ (apart from small corrections)
\begin{equation}
{\rm tan} \Theta _d = \sqrt{\frac{m_d}{m_s}}
\end{equation}
\[
{\rm tan} \Theta _u = \sqrt{\frac{m_u}{m_c}} \, .
\]
The experimental value for ${\rm tan} \, \Theta _u$ given by the ratio
$V_{ub} / V_{cb} $ is in agreement with the observed value for
$\left( m_u / m_c \right) ^{1/2} \approx 0.07$, but the errors for both
$\left( m_u / m_c \right) ^{1/2}$ and $V_{ub} / V_{cb}$ are the same
(about 25\%). Thus from the underlying texture no new information is
obtained.\\
\\
\hspace*{0.3cm} This is not true for the angle $\Theta _d$, whose
experimental value
is due to a large uncertainty.: $\Theta _d = 12.7^{\circ } \pm 3.8^{\circ }$.
If $\Theta _d$ is given indeed by the square root of the quark mass ratio
$m_d /m_s$, which is known to a high accuracy, we would know $\Theta _d$ and 
therefore all four parameters of the CKM matrix with high precision.\\
\\
\hspace*{0.3cm} As emphasized in ref.\ (5), the phase angle
$\varphi $ is very close to 90$^{\circ }$, implying that the LQ--triangle
and the
unitarity triangle are essentially rectangular triangles. In particular the
angle $\beta $ which is likely to be measured soon in the study of the
reaction $B^{\circ } \rightarrow J / \psi K^{\circ }_s$ is expected to be close
to $20 ^{\circ }$.\\
\\
\hspace*{0.3cm} It will be very interesting to see whether the angles
$\Theta _d$ and
$\Theta _u$ are indeed given by the square roots of the light quark mass
ration $m_d / m_s$ and $m_u /m_c$, which imply that the phase $\varphi$
is close to or exactly $90 ^{\circ }$. This would mean that the light quarks
play the most important r$\hat{\rm o}$le in the dynamics of flavor mixing and $CP$
violation and that a small window has been opened allowing the first view
accross the physics landscape beyond the mountain chain of the Standard
Model.\\
\\
{\bf REFERENCES}
\begin{enumerate}
\item[1.] N. Cabibbo, {\it Phys. Rev. Lett. \underline{10}} (1963) 531.

\item[2.] M. Kobayashi and T. Maskawa, {\it Prog. Theor. Phys. \underline{49}}
(1973) 652.

\item[3.] See, e.g., L. Maiani, {\it in: Proc. 1977 Int. Symp. on Lepton and
Photon}\\
\hspace*{0.5cm}{\it Interactions at High Energies (DESY, Hamburg}, (1977), 867;\\ 
L.L. Chau and W.Y. Keung, {\it Phys. Rev. Lett. \underline{53}} (1984) 1802;\\ 
H. Fritzsch, {\it Phys. Rev. \underline{D32}} (1985) 3058;\\ 
H. Harari and M. Leurer, {\it Phys. Lett. \underline{B181}} (1986) 123;\\
H. Fritzsch and J. Plankl, {\it Phys. Rev. \underline{D35}} (1987) 1732.

\item[4.] F.J. Gilman, Kleinknecht, and B. Renk, {\it Phys. Rev.
 \underline{D 54}} (1996) 94.

\item[5.] H. Fritzsch and Z.Z. Xing, {\it Phys.\ Lett. \underline{B353}}
(1995) 114. 

\item[6.] H. Fritzsch, {\it Nucl. Phys. \underline{B155}} (1979) 189.

\item[7.] S. Dimopoulos, L.J. Hall, and S. Raby, {\it Phys. Rev.
 \underline{D45}} (1992) 4192;\\ 
L.J. Hall and A. Rasin, Phys. Lett. \underline{B315} (1993) 164;\\
R. Barbieri, L.J. Hall, and A. Romanino, {\it Phys. Lett. \underline{B401}}
(1997) 47.

\item[8.] C. Jarlskog, {\it in CP Violation, edited by
C. Jarlskog}\\
\hspace*{0.5cm}{\it (World Scientific, 1989)}, p. 3.

\item[9.] H. Fritzsch, {\it Phys. Lett. \underline{B184} (1987)} 391.

\item[10.] H. Fritzsch, {\it Phys. Lett. \underline{B189} (1987)} 191.

\item[11.] L.J. Hall and S. Weinberg, {\it Phys. Rev. \underline{D48}} (1993) 979.

\item[12.] H. Lehmann, C. Newton, and T.T. Wu,
{\it Phys. Lett. \underline{B384}} (1996) 249.

\item[13.] M. Neubert, {\it Int. J. Mod. Phys. \underline{A11}} (1996) 4173.

\item[14.] R. Forty, talk given at the Second International
Conference on $B$ Physics and $CP$ Violation, Honolulu, Hawaii, March
24 - 27, 1997.

\item[15.] G.C. Branco, L. Lavoura, and F. Mota,
{\it Phys. Rev. \underline{D39}} (1989) 3443.

\item[16.] S. Dimopoulos, L.J. Hall, and S. Raby,
{\it Phys. Rev. Lett. \underline{68}} (1992) 1984; Phys.\\
R. Barbieri, L.J. Hall, and A. Romanino, Phys. Lett. \underline{B401}
(1997) 47.

\item[17.] H. Fritzsch and X. Xing, {\it Phys.\ Rev.\ \underline{D57}} (1998)
594--597

\item[18.] A. Rasin, Report No. hep-ph/9708216.

\item[19.] Particle Data Group, R.M. Barnett {\it et al.},
{\it Phys. Rev. \underline{D54}} (1996) 1.

\item[20.] H. Fritzsch, {\it Phys. Lett. \underline{B70}} (1977) 436;
\underline{B73} (1978) 317.

\item[21.] A.B. Carter and A.I. Sanda, {\it Phys. Rev. Lett. \underline{45}}
(1980) 952;\\ 
I.I. Bigi and A.I. Sanda, {\it Nucl. Phys. \underline{B193}} (1981) 85.

\item[22.] A.J. Buras, {\it Report No. MPI-PhT/96-111}; and
references therein.

\item[23.] L. Wolfenstein, Phys. {\it Rev. Lett. \underline{51}} (1983) 1945.

\item[24.] See, e.g., K.S. Babu and Q. Shafi, {\it Phys. Rev. \underline{D47}}
(1993) 5004;\\
\hspace*{0.5cm} and references therein.

\item[25.] H. Fritzsch and Z. Xing, in preparation.
\end{enumerate}
\end{document}